\documentclass[sigconf]{acmart}
\usepackage[utf8]{inputenc}
\usepackage{microtype}



\title{Virtual Fidgets: Opportunities and Design Principles for Bringing Fidgeting to Online Learning}
\author{Sam Ross}
\email{samhross@cs.washington.edu}
\affiliation{%
  \institution{University of Washington}
  \city{Seattle}
  \state{WA}
  \country{USA}
}
\author{Nicole Sullivan}
\email{nsulliv@cs.washington.edu}
\affiliation{%
  \institution{University of Washington}
  \city{Seattle}
  \state{WA}
  \country{USA}
}
\author{Jina Yoon}
\email{jinayoon@cs.washington.edu}
\affiliation{%
  \institution{University of Washington}
  \city{Seattle}
  \state{WA}
  \country{USA}
}
\date{December 2022}
\graphicspath{{images/}}

\copyrightyear{2023}
\acmYear{2023}
\setcopyright{rightsretained}
\acmConference[CHI EA '23]{Extended Abstracts of the 2023 CHI Conference on Human Factors in Computing Systems}{April 23--28, 2023}{Hamburg, Germany}
\acmBooktitle{Extended Abstracts of the 2023 CHI Conference on Human Factors in Computing Systems (CHI EA '23), April 23--28, 2023, Hamburg, Germany}\acmDOI{10.1145/3544549.3585729}
\acmISBN{978-1-4503-9422-2/23/04}

\begin{document}

\begin{abstract}
We present design guidelines for incorporating fidgeting into the virtual world as a tool for students in online lectures. Fidgeting is associated with increased attention and self-regulation, and has the potential to help students focus. Currently there are no fidgets, physical or virtual, designed for preserving attention specifically in online learning environments, and no heuristics for designing fidgets within this domain. We identify three virtual fidget proxies to serve as design probes for studying student experiences with virtual fidgeting. Through a study of eight students using our virtual fidget proxies in online lectures, we identify eight emergent themes that encompass student experience with virtual fidgeting in lectures. Based on these themes, we present four principles for designing domain-specific virtual fidgets for online lectures. We identify that virtual fidgets for lectures should be context-aware, visually appealing, easy to adopt, and physically interactive.
\end{abstract}

\maketitle

\section{Introduction}
Many people at some time or another feel an urge to fidget. In lectures, it is common to fiddle with objects in one’s surrounding environment. While this may seem like a distraction, fidgets afford people a way to release pent up energy \cite{levine}, allowing them to pay more attention and even increase retention of lecture material \cite{farley}. Most spontaneous fidget objects end up being non-distracting items around us, such as office supplies. In the virtual world, however, the landscape is sparse for potential virtual fidgets that are not designed to draw attention from the user. When a student watching a video lecture has trouble focusing, they are likely to turn to activities unrelated to the lecture, such as social media \cite{zureick} which, in the virtual world, is often the student’s closest “fidget object”. Social media, by its very nature, is designed to firmly hold attention \cite{bhargava}, and prior work has shown that students’ attention and retention significantly worsen when completing tasks on computers during lectures \cite{risko_everyday_2012, risko_everyday_2013}. These digital distractions have become even more common in the age of Zoom and online learning during the COVID-19 pandemic, with one large-scale study finding that 23-31\% of people reported multitasking during meetings \cite{cao_large_2021}. Trying to maintain attention during long video conferences can also lead to emotional burnout, or what has been commonly referred to as “Zoom fatigue” \cite{nesher_shoshan_understanding_2022}. 

In traditional classroom settings, fidgeting has particularly been shown to greatly improve children’s learning and attention \cite{carson_sit_2001, farley, grodner2015fidget}, which may have sparked the growth of fidget toys \cite{biel2017fidget}. While physical fidget tools do provide benefit to many people, as more classes move to an online or hybrid structure due to the COVID-19 pandemic, it is essential for students to have integrated ways to both maintain focus and bounce back from temporary inattention. We propose translating these fidgeting principles into virtual learning environments as a response to these changes. We believe virtual fidgets have a high potential for impact specifically in online lectures, where students often passively watch lectures without active engagement in the form of participation. 

This paper seeks to understand how students feel about using virtual fidgets on the same screen as their online lectures. We theorize that giving users a satisfying fidget located directly on the screen next to their lecture may fulfill their need for stimulation, keeping them more engaged in the lecture than if they were simultaneously using social media. To explore this, we designed a user study where students use virtual fidgets on the same screen as an online lecture. The insights from our users' perspectives were then used as a basis to provide recommendations for future designs. As a result of this work, we identify that the fidget intervention reduced transfer to other sites in a majority of participants. Based on our study, we present four core design principles to serve as guidelines for future virtual fidget tools. We suggest that virtual fidget tools should be context-aware, visually appealing, easy to adopt, and physically interactive.

\begin{figure*}[htp]
    \centering
    \includegraphics[width=\textwidth]{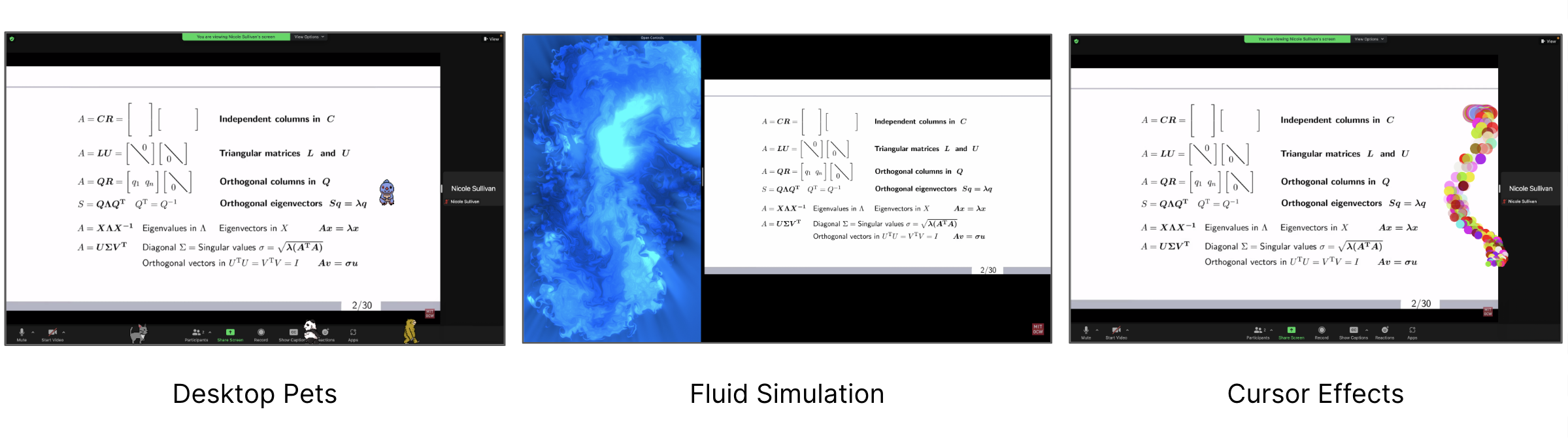}
    \caption{Images of the three fidget tools chosen for this study, set up in a Zoom lecture.}
    \label{fig:fidgets}
    \Description{Three screenshots labeled Desktop Pets, Fluid Simulation, and Cursor Effects. Each screenshot shows one of the three tools used in this study, set up in its respective Zoom environment. The Desktop Pets screenshot shows four small digital pets (dinosaur, cat, panda, and sloth) walking around the Zoom lecture screen. The Fluid Simulation screenshot shows a panel of blue abstract swirls to the left of the Zoom lecture. Finally, the Cursor Effects screenshot shows a trail of colorful medium-sized dots on the right side of the Zoom lecture screen.}
\end{figure*}

\section{Related Work}
Building on discoveries in psychology, HCI researchers have explored computing-supported artifacts to improve the experiences of online video conferencing through movement. Active workstations, for example, focus on physical activity while working \cite{choi2016exploring}, which addresses Bailenson’s suggestion that the reduced physical mobility during meetings may be a key contributor to Zoom fatigue \cite{bailenson_nonverbal_2021}. Others have created smartphone apps that utilize vibration for tactile, haptic feedback for the purpose of attention regulation \cite{williams2018user}. Research on mindfulness \cite{chin_mindfulness_2021} and affective experience on work have further inspired HCI systems work \cite{seo_role_2004}. A few examples include Haas \& An’s fashionable wearables that promote emotional regulation for people with anxiety \cite{haas_conceptual_2022}, Roquet et al.’s Purrrble teddy bears that embody emotional self-regulation for youth’s mental health \cite{dauden_roquet_exploring_2022}, and Ji \& Isbister's AR fidgets that place users in immersive experiences to regulate their mood \cite{ji_ar_2022}. Audience response systems are another alternative that can activate attention by adding an interactive component to a lecture. However, they are not initiated by the student, and past work has found that students can feel negatively about being monitored by such systems \cite{kay2009examining}. 

We find much of Karlesky \& Isbister’s work on “fidget widgets” useful in our investigation of virtual fidgets, and chose apps for our study inspired by their four key themes of fidget widgets: Tangential, Playful, Digital, and Tangible \cite{karlesky_fidget_2013}. Our investigation is unique in that although some of our apps do have tangible elements, we do not introduce new physical objects or devices in addition to the computers that participants already had access to, whereas prior work often incorporates custom hardware. We also adopt Karlesky et al.’s recommendation to think about designing for the “margins” of a workspace \cite{karlesky_understanding_2016}, motivated by the concept of spatial contiguity, which finds that placing instruction material close to each other helps with learning \cite{ginns_integrating_2006}. Chalkley et al. also found that physical movement---which prior research has established is correlated with better retention---can actually increase when screen engagement goes up \cite{chalkley2017wearable}. Additionally, considering that gazing away from screens is distracting and appears to be disrespectful \cite{marlow_taking_2016}, collocating it may also benefit fellow participants of the lesson. 

Our investigation includes the unique contribution of investigating virtual fidgets as an intervention for a specific task. Prior work has developed novel fidgets for self-regulation in both the physical and virtual world. Our study goes further by investigating the use of virtual fidgets for a particular goal, which in this case is to help students deal with distractions during online lectures. Our experiment design resembles da Câmara et al.’s work on identifying children’s fidget object preferences by using existing apps as design probes to elicit feedback about opportunities for future virtual fidgets \cite{da_camara_identifying_2018}. While fidgets have historically been associated with children, we differentiate our work by focusing specifically on adult college and graduate level students in the context of online video lectures. We also do not introduce new devices in the space, and collocate the fidget "in the margins" of the lecture material (on the screen).

\section{Methods}
This paper aims to provide a basis for incorporating a fidget into virtual learning spaces without requiring an external object. Thus, we did not seek to design the "ideal" virtual fidget, but rather explore guidelines for a future tool. For this investigation, we broadly consider virtual fidgets interactive tools where the primary input and feedback happen on the user's device. To do this, we had participants use pre-existing apps that have the potential to serve as virtual fidgets. In addition to drawing on Karlesky et al.'s four themes of "fidget widgets", we also selected the following three tools because they closely mimicked a “fidget object”: users can interact with the tool without it holding their direct attention for an extended period of time \cite{karlesky_fidget_2013}. The tools ultimately chosen for this study were: Desktop Pets \cite{curzel_2021}, Fluid Simulation \cite{dobryakov_2017}, and CursorEffect2 \cite{zhou_2021} (Figure \ref{fig:fidgets}). Desktop Pets is a macOS app which provides users with small, pixelated "pets" that walk around their screen in front of all other applications. Users can interact with the pets by clicking and dragging them with their mouse. Fluid Simulation is a web application that allows users to generate colorful swirls of light by clicking and dragging their mouse on a blank canvas. To use the Fluid Simulation, participants opened a split screen with both their virtual lecture and this fidget tool. CursorEffect2 is another macOS app that decorates users' cursors with animations like lights and colors as they move their mouse. Unlike the other two apps, CursorEffect2 does not require users to click with their pointing devices in order to activate its visual effects. 

We recruited eight participants at the University of Washington by publicizing the study on multiple community pages. All participants were students who had recently attended at least one online lecture. We presented participants with a Google Form that included all instructions and questions. Participants were first asked to choose one of three fidget tools based on what appealed to them. We then asked participants to watch an online lecture of their choosing, lasting at least thirty minutes, while using the tool. Afterwards, participants were tasked to answer several feedback questions. The questionnaire begins with the participants' experience with Zoom lectures in general. We used this as a baseline to determine if participants typically become distracted during online lectures. We also asked participants to specify what they usually do if they lose focus during virtual classes. After participants finished the lecture, participants recounted their experience using the fidget tool by answering a mix of Likert scale and free-response questions. 
 
All three authors then together performed a thematic analysis. We organized the open-ended responses by gathering themes that might inform the design of future virtual fidget tools. The questions were: "How did you feel about having the fidget on your screen along with the lecture?" "What were 2 things you liked about using the fidget?" "What were 2 things you disliked about using the fidget?" "What do you wish the fidget could do that it currently does not?". We utilized a collaborative cardsorting method to complete this thematic analysis. We printed out the individual responses on paper and and collaboratively created codes for recurring themes in the responses. Each time we felt a response matched an existing theme or warranted a new theme, we labeled it accordingly with the code. All three researchers had to agree on the code before marking the decision. 

\section{Results}
\subsection{Experiences with Distraction in Virtual Lectures}

All of the participants reported getting distracted during their most recent video lecture, and all reported using other sites during these moments of distraction, such as Facebook, Instagram, Twitter, Email, Slack, and Messenger/iMessage. These responses provided a baseline for the type of distraction common in video lectures. Each of these secondary sites involve reading text, which can make it difficult to simultaneously process the lecture itself. Similarly, many of these sites are designed to grab attention. A student with no alternative to watching their lecture may eventually find their way to focusing again simply from the lecture being the main stimulus. However, once attention has been transferred to an attention-grabbing alternative, it can be very difficult to get back on track. 

\subsection{Reactions to the Virtual Fidgets}
\begin{figure}[htp]
    \centering
    \includegraphics[width=\linewidth]{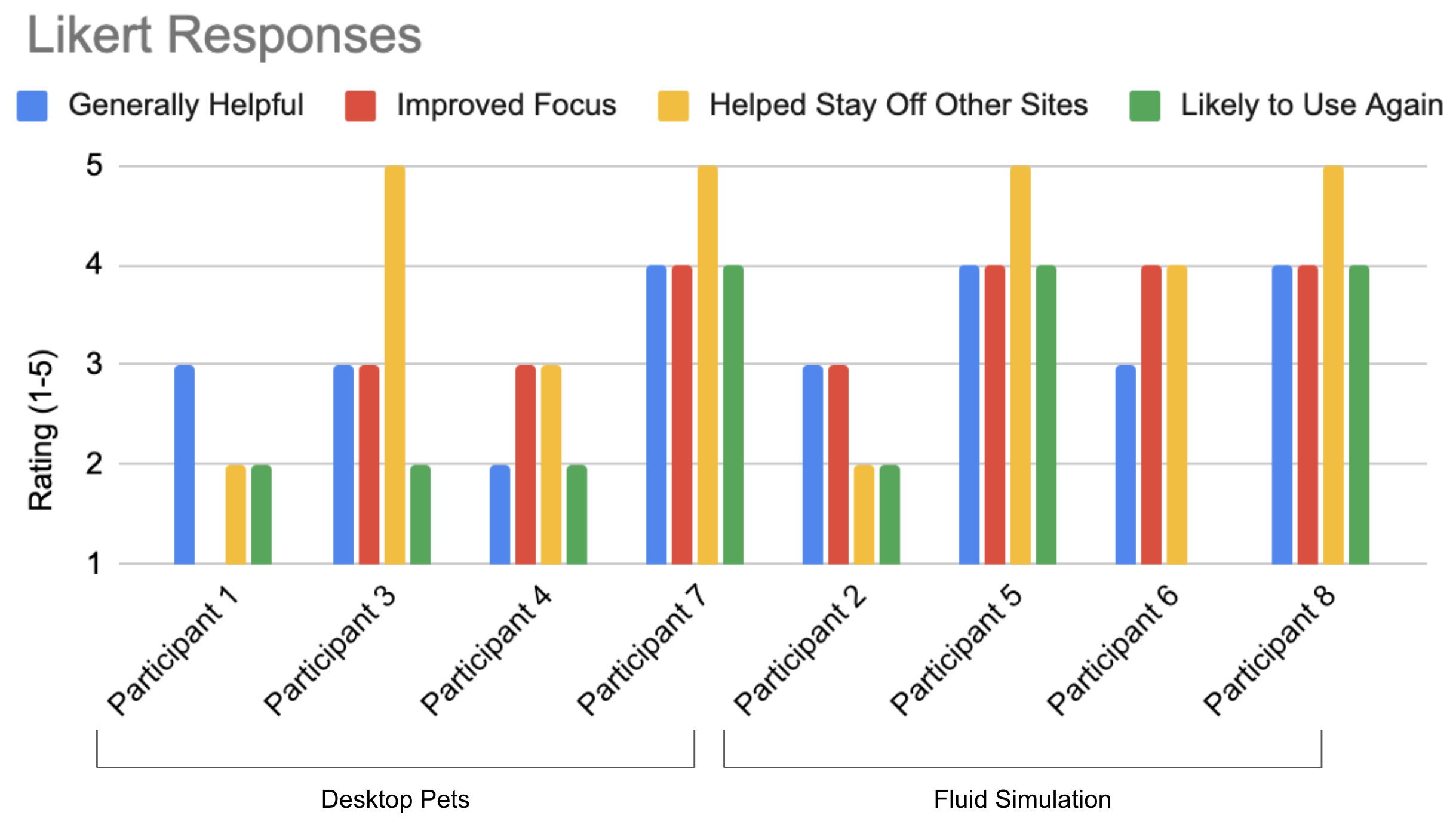}
    \caption{Participant ratings of virtual fidget effectiveness.}
    \label{fig:likert}
    \Description{Bar chart showing each of the eight participants' Likert scale responses. On the x-axis, each participant is associated with four bars, each bar representing the participant's answer to a question. Additionally, the participants are split between the the tool they used: Desktop Pets and Fluid Simulation. The summarized questions are (1) Generally Helpful, (2) Improved Focus, (3) Helped Stay Off Other Sites (4) Likely to Use again. The y-axis is the user's answer, which is a number from one to five. The highest scores were associated with the "Helped Stay Off Other Sites" question, in which four out of eight participants responded with a 5.}
\end{figure}

For the next section of the study, we prompted users to select one of three virtual fidget proxies based on what appealed to them. The goal of this section was to determine how users felt about our virtual fidget proxies, and to determine principles for future design of specialized virtual fidgets for online learning environments. All participants ended up choosing two of the possible three options, with half choosing Desktop Pets and half choosing the Fluid Simulation. Interestingly, no participants chose CursorEffect2, suggesting that the presentation of the app was not appealing to participants. Figure \ref{fig:likert} shows responses to each of the four Likert scale questions.

The biggest impact of the fidget was in preventing 62.5\% of users from switching to other sites. The other responses were mixed, with some indicating more usefulness and help with focusing than others. This indicates that there may be some individuals who benefit more from fidgeting than others, and that a tool designed for fidgeting should focus on those who perceive a benefit from fidgeting. Participants 5, 7, and 8 all rated they were likely to use the fidget again, indicating that despite the fact that these fidgets were not originally designed for fidgeting, they capture some of the necessary features of fidgets. 

The majority of data collected was in the form of short written responses, for which we conducted a thematic analysis. Through this activity, we identified eight themes that captured the range of participants’ thoughts about using the fidget with their virtual lecture, denoted by a designated letter code. The themes are presented in the following section with key examples of user feedback.

\subsection{Themes}
\textbf{Sites:} The user reported that the use of the fidget helped them not use other sites.
\begin{itemize}
    \item “I also think it is better than watching clips on Instagram or podcasts or people talking because that tends to sit in your subconscious a lot more and now you're just thinking about that. The Fluid Simulations don't affect my subconscious thoughts and so it's easy to get back to focusing on the lecture” (Participant 8, Fluid Simulation)
    \item “While I was expecting it will further distract me from the lecture, I am surprised that sometimes the pet prevented me from switching to another app and surfing other sites” (Participant 3, Desktop Pets)
\end{itemize}
\textbf{Physicality:} The user commented on the physical nature of the fidget.
\begin{itemize}
    \item “[It] satisfies the inner need to fidget or get distracted or play around with something in my hand…I could still look at it in my peripherals while moving the mouse around on the screen, and I would subconsciously be really satisfied while still focusing on the lecture.” (Participant 5, Fluid Simulation)
    \item “I will say I am more of a physical fidgeter since I still wanted to do something with my hands—i.e. fidget with my hair or spin my pencil, so I don't think the Desktop Pets really changed that urge.” (Participant 3, Desktop Pets)
\end{itemize}
\textbf{Novelty:} The user had a different experience of the fidget at first compared to once they were familiar with it. For example, some reported finding the fidget distracting at first, while others found the fidget repetitive over time.
\begin{itemize}
    \item “It was a little distracting at first, but not particularly distracting after the first 5 minutes.” (Participant 8, Fluid Simulation)
    \item “I didn’t like that the virtual pet moved back and forth in a repetitive straight line that became predictable” (Participant 7, Desktop Pets)
\end{itemize}
\textbf{Companionship:} The user commented on a feeling of having a companion due to the presence of the fidget.
\begin{itemize}
    \item “I liked it! Felt like the virtual pet was stuck on the Zoom with me” (Participant 7, Desktop Pets)
    \item “With the cat at my Zoom meeting, it feels like I had a pet alongside in my apartment” (Participant 4, Desktop Pets)
\end{itemize}
\textbf{Location:} The user referenced the location of the fidget in relation to their experience. For example, some comments positively enjoyed the proximity of the fidget to their lecture, while others encountered issues where the fidget got in the way of parts of the screen.
\begin{itemize}
    \item “I didn't like the detached interface, because I would like to have this as a background thing while I'm able to see the Zoom call in full screen” (Participant 8, Fluid Simulation)
    \item “Once the Desktop Pets were set up with a certain browser window structure (I wanted the pets to walk on top of the browser windows) I didn't want to change browsers—looking at the fidget when it framed the video player helped me look at the slides more instead of listening to the audio and scrolling on my phone which is what I usually do” (Participant 3, Desktop Pets)
\end{itemize}
\textbf{Intrusiveness:} The user described the intrusiveness of the fidget to their experience.
\begin{itemize}
    \item “It was non-intrusive, I could control when to use it” (Participant 6, Fluid Simulation)
    \item “I also didn’t like having to click and drag it out of the way when I wanted to see something specific” (Participant 7, Desktop Pets)
\end{itemize}
\textbf{Focus:} The user described a change in focus as a result of the fidget.
\begin{itemize}
    \item “Just the act of interacting with the fidget made me realize that I have to focus back on the lecture, although it was a fleeting realization, it at least helped me realize it quickly” (Participant 6, Fluid Simulation)
    \item “I enjoyed watching the pets walk around the screen. Whenever they walked close to the video player, I could feel my attention shift towards the slides that were being shared” (Participant 3, Desktop Pets)
\end{itemize}
\textbf{Visual:} The user commented on the aesthetic appeal of the fidget. 
\begin{itemize}
    \item “It was nice for a little break and cute to look at” (Participant 1, Desktop Pets)
    \item “It was a nice visual while watching a mainly static PowerPoint” (Participant 2, Fluid Simulation)
\end{itemize}

\begin{figure}[htp]
    \centering
    \includegraphics[width=\linewidth]{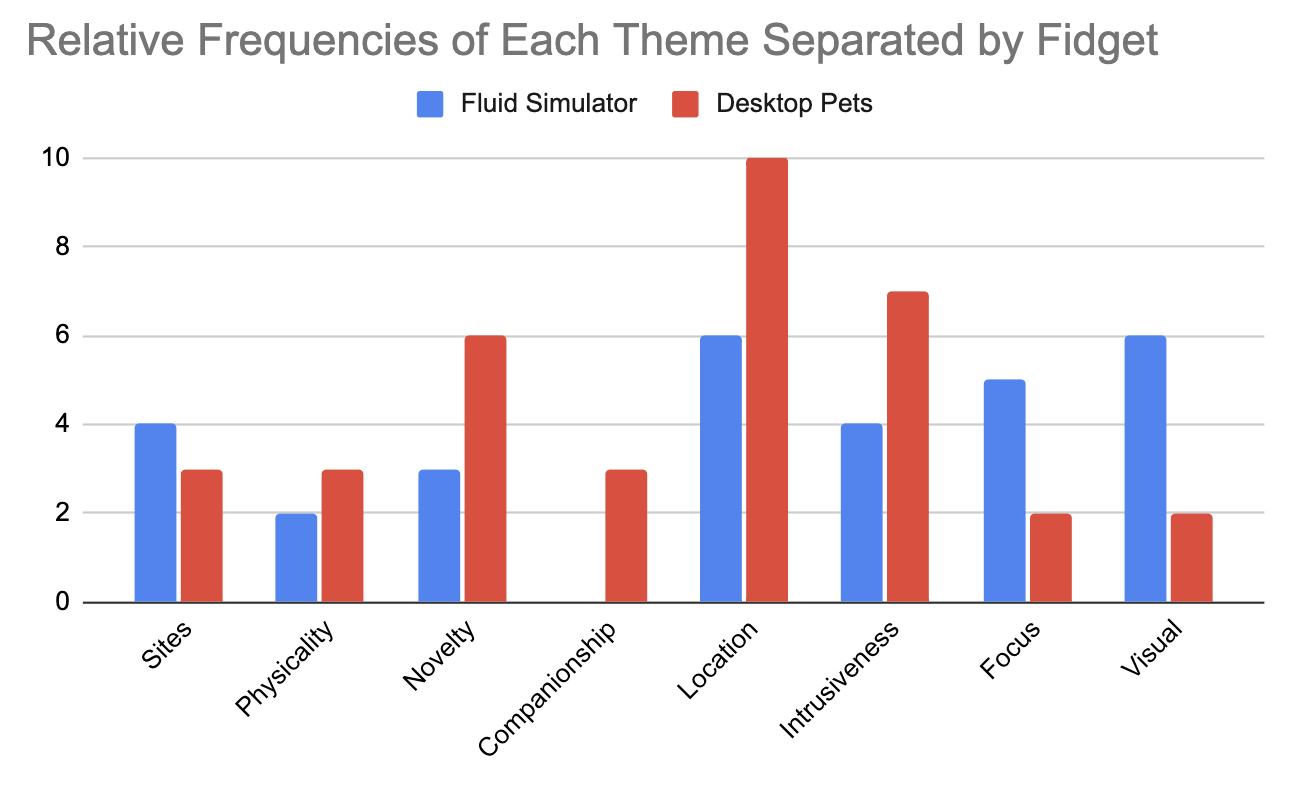}
    \caption{How often each theme occurred among the 32 responses.}
    \label{fig:themesbyfidget}
    \Description{Bar chart showing the relative frequency of each of the eight themes, separated by fidget. On the x-axis, each theme is associated with two bars. Each bar represents the number of themes found, separated by Fluid Simulation and Desktop Pets. The highest number of frequencies found are associated with Desktop Pets, in which ten counts were found for the Location theme, seven for the Intrusiveness theme, and six for the Novelty theme. The highest frequencies found for Fluid Simulation were six for the Location theme, six for the Intrusiveness theme, and five for the Visual theme.}
\end{figure}

Figure \ref{fig:themesbyfidget} shows the relative frequencies of each theme, separated by fidget used. Users most frequently brought up the location and intrusiveness of the fidget when discussing their experience, indicating that these factors have a large impact on the user experience. Some of these themes appeared more for one fidget than the other. Participants who used the Desktop Pets fidget brought up Location, Intrusiveness, and Novelty more often than those using the Fluid Simulation. The Desktop Pets fidget involved characters moving autonomously across the screen, and many people noted having to move them out of the way or having trouble figuring out how to use them. In contrast, the Fluid Simulation had more mentions of keeping people off other sites, as well as helping focus and being visually appealing. 

\section{Discussion}
The intervention of adding a virtual fidget in the learning environment was successful as an intermediary distraction barrier for a majority of participants. It gave students an outlet for distraction to prevent switching to other sites, often helping participants return to focus on the lecture. We found that students' impressions of the virtual fidgets were overall positive, and the majority of the issues students had with the fidgets stemmed from the fact that they were not designed for fidgeting or with lectures in mind. This suggests that tools specifically built for virtual fidgeting is a worthwhile intervention in online learning, where attention is vital and distraction via social media is common. 
\subsection{Principles for Designing Virtual Fidgets}
Based on our user feedback, we have identified the particular features of the Fluid Simulation and Desktop Pets fidget proxies that students liked and disliked. Our themed groupings show both areas of consideration in the design of fidgets, as well as fidgeting outcomes. From this feedback, we propose principles for designing domain-specific virtual fidgets intended for students watching online lectures.

\textbf{Principle 1: The fidget should be context-aware} \emph{Based on the themes of Location and Intrusiveness}. One of the most powerful interventions provided by the fidget was the fact that the fidget had the capacity to re-engage students by giving them an interaction close to the lecture itself. Equally importantly, the fidget had the capacity to interrupt the experience for the user if it obstructed important content. Therefore, designs of virtual fidgets should understand the interface in which they are used, and leverage that information to promote engagement. This can come in the form of points of interest created by the fidget that relate spatially to the content of the lecture. Designs should also avoid obscuring interface elements of the lecture that the user will always need to access. 

\textbf{Principle 2: The fidget should promote physical interaction} \emph{Based on the theme of Physicality}. Despite the predominantly visual nature of the fidgets, part of the satisfaction students reported relied on the physical interaction between the student and their mouse. It was not enough for students to have an engaging visual among their lecture; it was important that the act of fidgeting also included motor input. This is consistent with traditional fidgeting, where physical movement helps release pent-up energy. Therefore, designs of virtual fidgets should consider how the user interacts with the fidget, and consider how standard input devices can function as satisfying fidget inputs. 

\textbf{Principle 3: The fidget should be visually appealing} \emph{Based on the themes of Visuality, Focus, Companionship, and Other Sites}. 
Given that participants reported frequently navigating away from lectures, the virtual fidget should be designed with enough visual appeal to keep their attention. The fidget must be designed to capture the interest of the user, while not being so distracting as to pull all focus from the lecture. The visuals of the fidget also help keep the attention of the user close to the lecture content. Traditional fidget tools are always available but often engage the user both physically and visually, requiring the user to look away from the screen and toward the object. In contrast, virtual fidgets can be manipulated with the hands and provide visual feedback at the desired point of attention on the screen. Maintaining the visual interest of the virtual feedback gives users both an outlet to fidget and an extra reason to look where they are supposed to for their lecture.

\textbf{Principle 4: The fidget should be easy to adopt} \emph{Based on the theme of Novelty}. Many users reported initial confusion while figuring out how to use the fidget. For example, Desktop Pets already has features that allow users to pick up and move the pets, yet several participants said they wished they could move the pets around. This implies that the interactions were not easily discoverable. Both fidgets had various settings the user could manipulate which was a distraction for many participants. Therefore, applications designed for fidgeting should be simple and self-explanatory. 

\section{Limitations}
An important but intentional limitation is that our findings are not generalizable. This is because the study was conducted on a small sample of students, and our survey did not collect data about their demographic backgrounds. We also recognize that fidgeting is not universally helpful. Similar to how Homer et al. \cite{homer_effects_2008} find that the effects of visual educational tools depend on individuals’ pre-existing preferences for visual learning, the effects of fidgeting likely depend on individuals’ pre-existing preferences for fidgeting. This is consistent with suggestions that those with ASD or ADHD are more likely to benefit from physical fidgets than others \cite{biel2017fidget}. Given our application in online learning, it would be valuable to investigate whether these virtual fidgets are helpful to those with learning disabilities. It is important to note that although we point to this literature as an example of variations in preferences, we do not make claims that virtual fidgets necessarily benefit these specific groups as we did not design this study with such populations \cite{spiel2022adhd}.

The context of our study is also limited in scope because, as intended, it was designed specifically for online lecture environments, so our findings may only apply to educational settings that do not involve interaction between participants (e.g. discussion-based classes or conference meetings). Additionally, the apps we chose were used as design probes or proxies and have their own limitations as they were not developed for the purpose of virtual fidgets. The novelty effect of these apps could also diminish over time. Lastly, all data in this study were self-reported, so we do not make claims about quantifiable changes to attention, and instead focus on how these virtual fidgets improved participants' affective experience.

\section{Future Work}
This investigation is a preliminary step in understanding the role of virtual fidgets in online learning and provides a basis for future work on the subject. In particular, because some responses showed particular enthusiasm for the virtual fidgets over others, future work could engage more deeply with those who benefit most from fidgeting and their needs in virtual learning environments. We recommend leveraging user-centered and/or participatory design methods to ensure it is designed well with and for a more specific group of intended users. If future work implements such a tool, a user evaluation may also be beneficial to investigate whether participants’ self-reports are consistent with measurable outcomes. Future work should also investigate the optimal level of stimulation a virtual fidget should offer to balance being engaging without being too distracting, which may vary between individuals. Additionally, this project has investigated virtual fidgets in online lectures only from the perspective of students. Future work could look at how instructors feel about their students using virtual fidgets, especially because instructors may perceive the fidgets as distracting. There are elements of social presence and perception of a virtual fidget that could be investigated, such as an indicator when a student is using a virtual fidget to avoid misunderstandings or incorrect assumptions \cite{marlow_taking_2016}. This would also reduce the virtual fidget user’s stress of monitoring self-presentation \cite{george2022users}. Alternatively, instructors might see virtual fidgets as an opportunity to engage students in a new way, and future work could develop fidgets that are responsive and relevant to their specific lecture content. We envision that the tool may even be adapted to be used as a way to deploy teaching material, like Schacter and Szpunar's \cite{schacter_enhancing_2015} work on integrating regular quizzes or comprehension checks, or a lecture-related visual aid \cite{johnson_eye_2012}. Finally, lectures are not the only times people feel a need to fidget, and future work could consider other contexts in which virtual fidgeting is appropriate and beneficial, such as pre-recorded videos or interactive meetings. Informing the design of features to support attention in video meetings with Kuzminykh and Rintel’s \cite{kuzminykh_classification_2020} functional recommendations for video conferencing UI would also be useful when implementing a virtual fidget that lives directly on or next to the Zoom interface. 

\section{Conclusion}
This study was an initial investigation on how students may feel about virtual fidgets in online educational lecture environments to help with attention regulation. In addition to creating an enjoyable experience, we hope this work helps ensure that the global shift toward remote work and learning is inclusive to those who may be more susceptible to distractions online. In summary, our work contributes eight major themes extracted from participants’ responses and four key design principles based on our results. We have shown that virtual fidgets can benefit certain students for the specific task of watching online lectures and that the concept is worth pursuing in future work.

\begin{acks}

Thanks to James Fogarty and Lisa Elkin for their guidance with this project. This material is based upon work supported by the NSF Graduate Research Fellowship under Grant No. DGE-2140004. Any opinions, findings, and conclusions or recommendations expressed in this material are those of the authors and do not necessarily reflect the views of the National Science Foundation.
\end{acks}

\bibliographystyle{ACM-Reference-Format}
\bibliography{citations}

\end{document}